\documentclass[a4paper,twocolumn,prl,showpacs]{revtex4}
\usepackage{amssymb}
\usepackage{graphicx}

\begin{document}

    \title{Life, Death and Preferential Attachment}
    \date{\today}

    \author{S. Lehmann}
    \affiliation{Informatics and Mathematical Modeling. \\
    Technical University of Denmark, Building 321, DK-2800 Lyngby, Denmark}
    \author{A. D. Jackson}
    \affiliation{The Niels Bohr Institute. \\
    Blegdamsvej 17, DK-2100 Copenhagen \O, Denmark}
    \author{B. Lautrup}
    \affiliation{The Niels Bohr Institute. \\
    Blegdamsvej 17, DK-2100 Copenhagen \O, Denmark}


\begin{abstract}
Scientific communities are characterized by strong stratification.
The highly skewed frequency distribution of citations of published
scientific papers suggests a relatively small number of active,
cited papers embedded in a sea of inactive and uncited papers.  We
propose an analytically soluble model which allows for the death
of nodes.  This model provides an excellent description of the
citation distributions for live and dead papers in the SPIRES
database.  Further, this model suggests a novel and general
mechanism for the generation of power law distributions in
networks whenever the fraction of active nodes is small.
\end{abstract}

\maketitle

That progress in science is driven by a few great contributions
becomes disturbingly clear when one considers citation statistics.
The vast majority of scientific papers is either completely
unnoticed or minimally cited. In high energy physics, $4\%$ of all
papers account for $50\%$ of the citations, while $29\%$ of all
papers are not cited at all~\cite{lehmann:03}.

In a pioneering sociological work analyzing American high energy
physicists, Cole and Cole~\cite{cole:73} connect this high degree
of stratification in the scientific literature to what they call
\emph{cumulative advantage}. The concept underlying cumulative
advantage was originally introduced by
R.~K.~Merton~\cite{merton:68} with the more striking name of the
`\emph{Matthew Effect}'. Merton's simple observation was that
success seems to breed success. A paper which has been cited many
times is more likely to be cited again than one which is less
cited, since ``unto every one that hath shall be given, and he
shall have abundance: but from him that hath not shall be taken
away even that which he hath''\cite{bible:03}---hence the name.

Inspired by Refs.~\cite{cole:73,merton:68} and his own work on
citation networks~\cite{price:65}, de Solla Price recast
Simon's~\cite{simon:57} ideas on the mathematics leading to the
power law distributions found in nature and society into the first
mathematical model of a scale-free network~\cite{price:76}. Much
later, the principles underlying Price's model were independently
re-discovered by Barab\'asi and Albert~\cite{barabasi:99}, who
coined yet another name for the same effect, namely
\emph{preferential attachment}. Preferential attachment has since
become a widely accepted explanation of the power law degree
distributions in complex networks in general. The strength of the
preferential attachment model in either incarnation is its
simpli\-ci\-ty, but this can also be its weakness.  In particular,
such models tend to assume that networks are homogeneous.  When
real world networks can be shown to have identifiable and
significant inhomogeneities, preferential attachment must be
supplemented by appropriate additional ingredients.

For example, it is an empirical fact that the vast majority of
nodes in citation networks ``die'' after a relatively short time
and are never cited again.  A relatively small population of
papers remains alive and continues to accumulate citations many
years after publication; this is the main conclusion in
Ref.~\cite{lehmann:03}.  The distinction between live and dead
populations represents an important inhomogeneity in the citation
data that is not considered in the simple preferential attachment
model. We do not suggest that the presence of death in citation
networks diminishes the importance of preferential attachment,
however, the distinctly different citation distributions observed
for live and dead papers compel us to include the effects of the
death of papers in our modeling efforts. It is the purpose of this
paper to suggest one such extension of preferential attachment
models.

\section{Dead Papers}\label{sec:data}
The work in this paper is based on data obtained from the SPIRES
database of papers in high energy physics. To be specific, the
data used below is the network of all citable papers from the
Theory subfield of SPIRES, ultimo October $2003$.  Filtering out
all papers for which no information of publication time is
available, we are left with a network of $275\,665$ nodes (i.e.,
papers).  All citations to papers not in this network were
removed, resulting in $3\,434\,175$ edges (i.e., citations).

\begin{figure}[htbf]
\centering
  \begin{tabular}{cc}
  \includegraphics[width=\hsize]{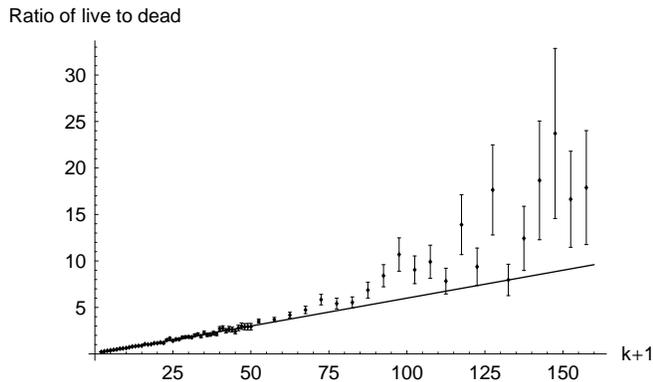}
  \end{tabular}
  \caption{\emph{The ratio of live to dead papers. The solid straight line has
  been inserted to illustrate the linear relationship between the
  live and dead populations for low values of $k$. The error bars are calculated
  from the square roots of the citation counts.}
  }\label{fig:invratio}
\end{figure}

Clearly, there is a variety of ways to define what is meant by a
dead node in real data\footnote{We recognize that there are
examples of papers that receive new citations after a long dormant
period.  However, such cases are rare and do not affect the large
scale statistics.}. We have tested several definitions, and our
results are qualitatively independent of the specifics of the
definition.  We have chosen to define papers that have not been
cited in 2003 to be dead.  Having identified a population of dead
papers, we have determined the citation distributions for live and
dead papers.  These distributions are shown in
Figure~\ref{fig:modelfit}(a) and indicate that the two
distributions are significantly different. As suggested in the
introduction, most (i.e., approximately three-quarters) of the
papers in SPIRES are dead.  It is also a simple matter to
determine the empirical ratio of live to dead papers as a function
of the number of citations per paper $k$.
Figure~\ref{fig:invratio}
displays this ratio in the range $1\le k \le 150$.  Over most of
this range the data is described by a straight line.  We note that
the data for dead papers with high $k$-values is very sparse.
Since only 0.15\% of dead papers have more than 100 citations,
statistics beyond this point are highly unreliable.  Thus,
plotting the ratio of live to dead papers gives a pessimistic
representation of the data. The ratio of dead to live papers is
described satisfactorily by the simple form $b/(k+1)$ for all but
the highest values of $k$, where this form overestimates the
number of dead papers by a factor of two to three. In short,
Figure~\ref{fig:invratio} implies that---to a fairly good
approximation---the fraction of dead papers with $k$ citations is
proportional to $1/(k+1)$.  We will make use of this fact in the
next section to suggest an extension of the preferential
attachment model which includes the effects of death.

\section{Modeling Death and Preferential Attachment}

Following the usual structure of preferential attachment models,
we imagine that at every update a new paper makes $m$ references
to papers already in the network and then enters the network with
$k = 0$ real citations and $k_0 = 1$ ``ghost'' citations. Since we
have chosen to eliminate all references to papers not in SPIRES in
constructing our data set, there is an obvious and rigorous sum
rule that the average number of citations per paper is also $m$.
The probability that a paper in the network will receive one of
these references is assumed to be proportional to its current
total of real and ghost citations. We can estimate when the
effects of preferential attachment become important by regarding
$k_0$ as a free parameter. Since we see no a priori reason why a
paper with $2$ citations should have a significant advantage in
acquiring citations over a paper with $1$ citation, we prefer to
allow the data to decide. Thus, in our model, the probability that
a paper with $k$ citations acquires a new citation at each time
step is proportional to $k+k_0$ with $k_0 > 0$.  We can think of
the displacement, $k_0$, as offering a way to interpolate between
full preferential attachment ($k_0=1$) and no preferential
attachment ($k_0\rightarrow\infty$).

More importantly, at every update each live paper in the network
has some probability of dying.  Guided by the SPIRES data, we
assume that this probability is proportional to $1/(k+1)$ for a
paper with $k$ real citations.  Once dead, a paper can no longer
receive new citations. In his 1976 paper, Price notes that
cumulative advantage is only half the Matthew Effect, because
although success is rewarded, there is no punishment for failure.
In this sense, the model described here represents one
implementation of the \emph{full} Matthew Effect.  Since the rate
at which papers are killed is inversely proportional to the number
of citations which they have, low cited papers have a much higher
probability of paying the ultimate penalty.

The rate equation approach introduced in the context of networks
by Krapivsky, Redner, and Leyvraz~\cite{krapivsky:00a} can easily
be modified to allow for death. We let $L_k$ be the probability
for finding a live paper with $k$ citations and $D_k$ be the
probability of finding a dead paper with $k$ citations. Each paper
cites $m$ other papers in the database. Papers are loaded into the
database with in-degree $k=0$. We arrive at the following rate
equations
\begin{eqnarray}
 L_k &=& m(\lambda_{k-1}L_{k-1}-\lambda_kL_k) - \eta_kL_k +
 \delta_{k,0}\label{eq:rateeqns1}\\
 D_k &=& \eta_k L_k,\label{eq:rateeqns2}
\end{eqnarray}
where $\lambda_k$ and $\eta_k$ are rate constants. We define $L_k$
to be equal to zero for $k<0$ and since every paper has a finite
number of citations, the probabilities $L_k$ must become exactly
zero for sufficiently large $k$. Thus, we can let all sums run
from $k=0$ to infinity. While the total citation distribution is,
of course, given by $L_k+D_k$, we can also probe the live and dead
distributions separately both theoretically and empirically. For
any choice of $\lambda_k$ and $\eta_k$ these equations trivially
satisfy the normalization condition on the total distribution.
However, the constraint that the mean number of references equals
the mean number of citations, $\sum_kk(L_k+D_k)=m$, must be
imposed by an overall scaling of the $\lambda_k$ and $\eta_k$.
Eq.~(2) shows that the coefficients, $\eta_k$, are simply the
ratio of dead to live papers as a function of $k$.  Given the
empirical values of this ratio shown in Figure 1, our model
corresponds to the case where
\begin{equation}\label{eq:etalambda}
    m \lambda_k=a(k+k_0) \quad \textrm{and}\quad
    \eta_k=\frac{b}{k+1}.
\end{equation}
Performing the recursion, we find
\begin{equation}\label{eq:solution}
  L_k = \frac{\Gamma(k+2)}{ak_1k_2}
  \frac{\Gamma(k+k_0)}{\Gamma(k_0)}
  \frac{\Gamma(1-k_1)}{\Gamma(k-k_1+1)}
  \frac{\Gamma(1-k_2)}{\Gamma(k-k_2+1)},
\end{equation}
where $k_1$ and $k_2$ are the solutions to the quadratic equation
\begin{equation}\label{eq:quadratic}
    (a(k+k_0)+1)(k+1)+b = 0
\end{equation}
regarded as a function of $k$.

One general observation of some interest emerges in the limit $k_0 \to
\infty$ in which preferential attachment is turned off.  We obtain
this limit by making the replacement $\alpha = a k_0$ in
Eq.~(\ref{eq:solution}) and then taking the limit $k_0\rightarrow \infty$
for fixed $\alpha$. A little work reveals that
\begin{equation}\label{eq:specsol}
    L_k=\frac{1}{\alpha}
    \left(\frac{\alpha}{1+\alpha}\right)^{k+1}
    \frac{(\frac{b}{1+\alpha})!(k+1)!}{(\frac{b}{1+\alpha}+k+1)!}
\end{equation}
The $D_k$ are simply $bL_k/(k+1)$ as before.
(Eq.~(\ref{eq:specsol}) can also be obtained by solving
Eqs.~(\ref{eq:rateeqns1}) and~(\ref{eq:rateeqns2}) with constant
$\lambda_k$ and $\eta_k = b/(k+1)$; the two approaches are
equivalent.)  When the death mechanism is eliminated by setting $b
= 0$, the resulting distribution shows an exponential decrease
which is to be expected given the assumed absence of preferential
attachment.

In fact, the death of nodes offers an alternative mechanism for
obtaining power laws.  To see this, consider the limit $\alpha \to
\infty$ and $b \to \infty$ with the ratio $r = b/(\alpha +1)
\approx b/\alpha$ fixed.  In this limit it is tempting to replace
the term $\alpha/(1+\alpha)$ by $1$, which allows us to compute
simple expressions for the fraction of dead papers $f$ and the
average number of citations of the live and dead papers, $m_L$ and
$m_D$.  (This approximation is appropriate when $r \ge 2$.  When
$r < 2$ the neglected factor is essential for ensuring the
convergence of $m_L$ and/or $m_D$.) The fraction of live papers is
then
\begin{equation}
1-f=\frac{1}{\alpha (r-1)},
\end{equation}
and the average number of citations for the live papers and dead papers,
respectively, is
\begin{equation}
m_L = \frac{2}{r-2} \ \ {\rm and}\ \ m_D = \frac{1}{r-1}.
\end{equation}
The average number of citations for all papers is evidently $m_D$
in the limit $\alpha \to \infty$ for which $f \to 1$.  Most
importantly, we see in this limit that
 \begin{equation}\label{eq:simplepowers}
 L_k \sim \frac{1}{k^r} \ \
{\rm and}\ \ D_k \sim \frac{b}{k^{r+1}}
\end{equation}
for $k > r$.  Thus, we see that power-law distributions for both
live and dead papers emerge naturally in the limit where the
fraction of dead papers $f$ goes to $1$. In this limit, a
vanishing fraction of live papers swim in a sea of dead papers.
Since such power laws are sometimes regarded as an indication of
preferential attachment, it is useful to see a quite different way
of obtaining them.

\section{Death in the Real World}
\begin{figure}[htbf]
  \centering
  \begin{tabular}{c}
  \includegraphics[width=\hsize]{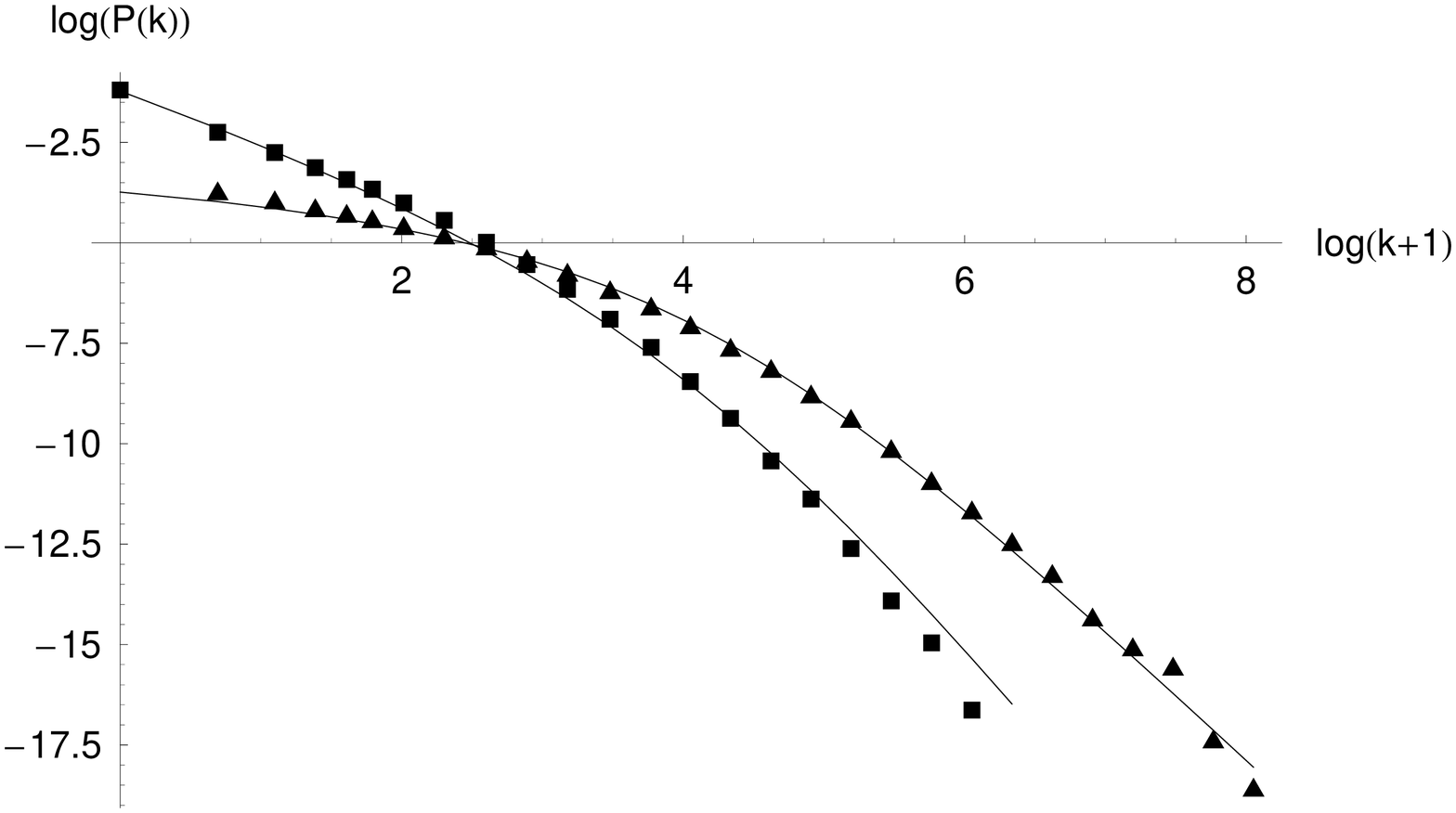}\\
  (a)\\
  \includegraphics[width=\hsize]{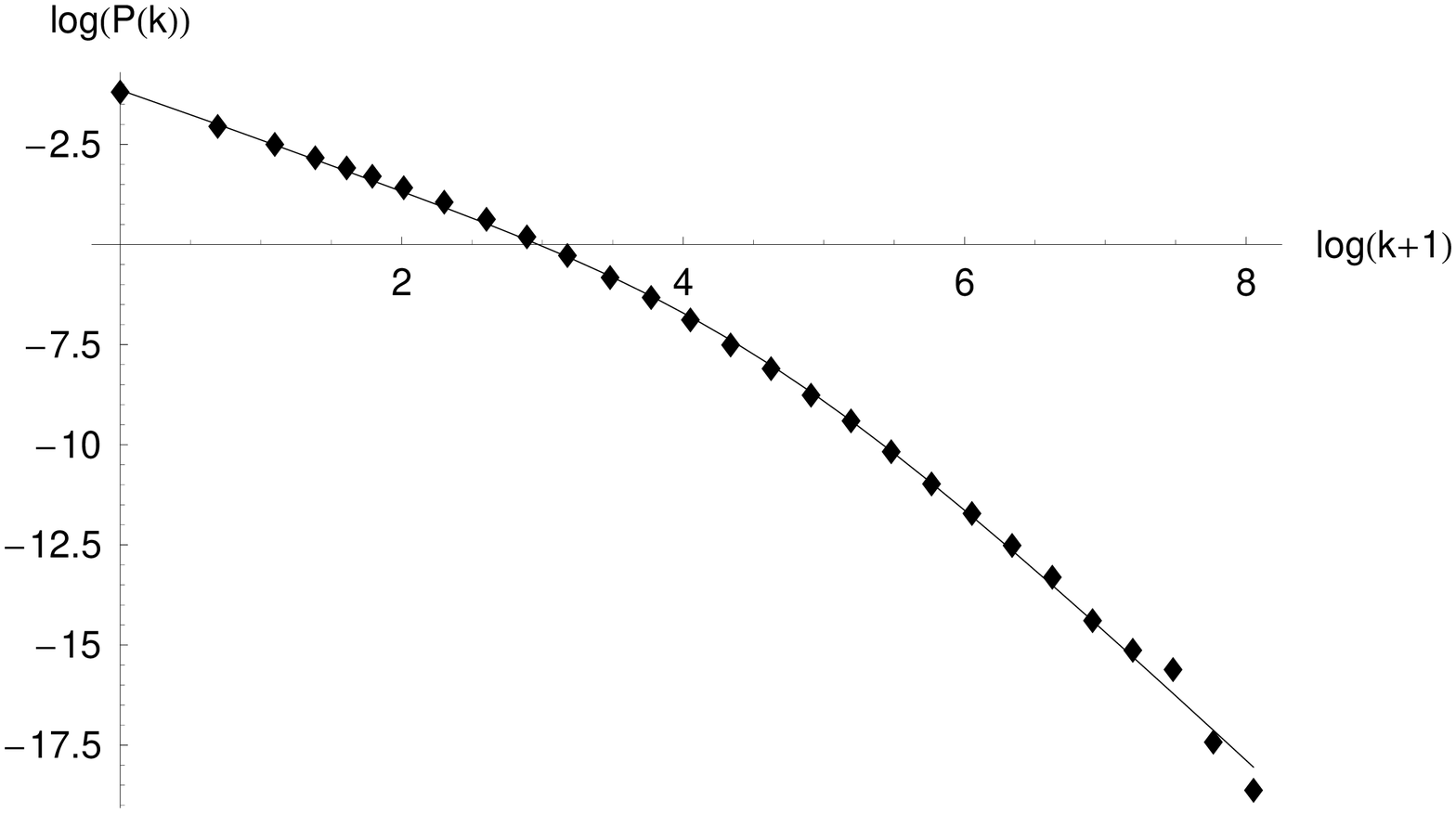}\\
  (b)\\
  \end{tabular}
  \caption{\emph{(a) Log-log plots of the distributions for live and
  dead papers. The triangles are the live data and the squares
  are the dead data. The solid lines are the fit.
  (b) A log-log plot of the distribution of all papers (live plus dead). The
  points are the data; the solid lines are the fit.}}\label{fig:modelfit}
\end{figure}

We now return to the full model and compare it to the data from
SPIRES. If we assign all zero cited papers to the dead category,
the mean number of citations is $34.1$ for live papers, $4.5$ for
dead papers, and 12.5 for all papers. The fraction of live papers
is $27.0\%$. By minimizing the squared fractional error, we can
fit the live data with an rms error of only $21\%$ using the forms
of Eqns.~(\ref{eq:solution}) and~(\ref{eq:quadratic}) with the
parameters $k_0=65.6$, $a=0.436$, and $b=12.4$. Given that the
data spans six orders of magnitude, the quality of this agreement
is strikingly high. The results of the fits are displayed in
Figure~\ref{fig:modelfit}.

The fitted mean number of citations is $32.9$ citations for live
papers, $4.25$ for dead papers, and $12.8$ for all papers.
According to the fit, $7.5\%$ of all papers with $0$ citations
are, in fact, alive. Assigning this fraction of zero citation
papers to the live data, we find mean citations of $31.5$, $4.6$,
and $12.5$ respectively. We also find that $29.2\%$ of the papers
in the model are live. This is in excellent agreement with the
data. There is remarkably little strain in the fit. We can, for
example, determine the model parameters $a$, $b$, and $k_0$ from
the empirical values of $m_L$, $m_D$, and $f$. This leads to small
changes in the model parameters and yields a description of
comparable quality for the distributions.  It is clear from
Figure~\ref{fig:modelfit} that the present fit to the live
distribution leads to some systematic errors in the description of
the dead population for the highest values of $k$. Given the
deviations from a straight line of the data of
Figure~\ref{fig:invratio} for large $k$, this comes as no
surprise. This could obviously be remedied by a small modification
of the $\eta_k$ through the inclusion of a suitable $k^2$ term in
the denominator.

It is clear that the present simple model is capable of fitting
the distributions of both live and dead papers with remarkable
accuracy. We note that the best fit value of the parameter $k_0 =
65.6$ suggests that a paper with $k = 66$ citations has a
competitive advantage over a paper with no citations of a factor
of $2$ rather than the factor of $67$ suggested by the simplest
preferential attachment models.

\section{Discussion and Conclusions}
It is obvious that the death mechanism introduced here is
essential if we wish to consider the empirical citation
distributions of live and dead papers separately.  It is less
obvious that the death mechanism (i.e., $b \ne 0$) is required to
provide a good description of the total citation data.  A similar
fit to the citation distribution for all papers with the
constraint $b=0$ yields the parameters $a=0.528$ and $k_0 = 13.22$
and gives an rms fractional error of $33.6$\%.  Although there are
some indications of systematic deviations in the resulting fit,
its overall quality remains high in spite of the fact that this
constrained fit ignores important correlations present in the data
set.  This result illustrates the familiar fact that more detailed
modeling is not necessarily required to fit global network
distributions even if important empirical correlations are
neglected in the process.  It also reminds us of the equally
familiar corollary that even a high quality fit to global network
distributions cannot safely be regarded as an indication of the
absence of additional correlations in the data. The most
significant difference between the model parameters obtained with
and without the death mechanism is the value of $k_0$, which
changes by a factor of $5$ from $65.6$ to $13.2$.  We have an
intuitive preference for the larger value.  (We believe that
preferential attachment will play an important role when a paper
is sufficiently visible that authors feel entitled to cite it
without reading it and that $k_0 \approx 65$ represents a
reasonable threshold of visibility.)  It is clear, independent of
such subjective preferences, that it is dangerous to assign
physical significance to even the most physically motivated
parameters if a network contains unidentified correlations or if
known correlations are neglected in the modeling process.
Specifically, it is difficult to draw firm conclusions regarding
the onset of preferential attachment if the death mechanism is not
included.

We have identified significant differences between the citation
distributions of live and dead papers in the SPIRES data, and we
have constructed a model including both modified preferential
attachment and the death of nodes that is quantitatively
successful in describing these differences.  We have further seen
that the death mechanism can provide an alternate mechanism for
producing power law distributions when the fraction of live nodes
is small.  Since many networks involve a small fraction of active
nodes, this mechanism may be of more general utility.  However,
the numerical success of the present model does not indicate the
absence of additional correlations in the SPIRES data.  In fact,
we know that such correlations exist. Consider the conditional
probability, $P(k | {\bar m})$, that a paper written by an author
with a lifetime average of ${\bar m}$ citations per paper will
receive $k$ citations.  The general interest in citation data is
based on the wide-spread intuitive belief that $P(k | {\bar m})$
is a sensitive function of ${\bar m}$.  This belief is supported
by the SPIRES data and will be treated in a subsequent
publication.

\begin{acknowledgments}
Our grateful thanks to Travis C.~Brooks at SPIRES without whose
swift replies and thoughtful help we would have lacked all of the
data!
\end{acknowledgments}

\bibliography{bibliography}
\bibliographystyle{unsrt}
\end{document}